\documentclass[journal]{IEEEtran}

\usepackage{pifont}
\usepackage{eurosym}
\usepackage{amsmath}
\usepackage{amssymb}
\usepackage{mathtools}
\usepackage[utf8]{inputenc}
\usepackage[vietnamese,english]{babel}

\usepackage{cite}

\ifCLASSINFOpdf
\usepackage[pdftex]{graphicx}
\else
\fi

\usepackage{amsmath}

\usepackage{url}

\usepackage{xcolor}
\usepackage{makecell}
\usepackage[subtle]{savetrees}

\begin{document}
\title{Sustaining Dynamic Traffic in Dense Urban Areas with High Altitude Platform Stations (HAPS)}

\author{Cihan~Emre~Kement, Ferdi~Kara, Wael~Jaafar, Halim Yanikomeroglu, Gamini Senarath, Ngọc Dũng Đào, and Peiying Zhu
}

\renewcommand{\baselinestretch}{1.1}
\selectfont 

\markboth{IEEE Communications Magazine}%
{Kement \MakeLowercase{\textit{et al.}}: Sustaining Dynamic Traffic in Dense Urban Areas with High Altitude Platform Stations (HAPS)}
%



\maketitle

\begin{abstract}
The impact of information and communication technologies on global energy consumption is increasing every year, and mobile networks account for a significant portion of it. More than 50\% of the total energy consumption of mobile networks is issued from radio access networks (RANs), due mainly to the rapid penetration of data-intensive applications and the increasing heterogeneity, dynamicity, and unpredictability of traffic. To tackle these high-demanding problems, RAN densification through the installation of additional base stations in high-demand areas is conventionally used. However, this leads to inefficient  energy use and over-provisioning issues.
In this context, high altitude platform stations (HAPS)
may be used to complement RANs and sustain their services in densely populated areas, where traffic can peak unpredictably. Due to their wide coverage areas, substantial communication payloads, and green energy model, HAPS super macro base stations (SMBSs) are capable of handling the massive and dynamic mobile data traffic of ground users. In this paper, we show how HAPS-SMBSs can complement RANs and serve the dynamic and unpredictable traffic demands of users in an energy-efficient manner. Through the simulation of a case study, we demonstrate the performance of a HAPS-SMBS compared to the conventional RAN densification method and analyze the two approaches in terms of sustainability.

\end{abstract}


%
\IEEEpeerreviewmaketitle

\section{Introduction}
%
%
%
%

To mitigate climate change and its impacts, many countries and organizations, such as the United Nations, are aiming to reach net zero CO$_2$ emissions by 2050 \cite{unitednations}. To achieve this goal, CO$_2$ emissions must be reduced by 45\% from their 2010 reference level within the next eight years, which means immediate action is needed in the design of future technologies. This is of paramount importance since the information and communication technologies (ICT) sector is expected to account for up to 20\% of global energy consumption by 2030 \cite{nextgen}.

In telecommunication networks, radio access networks (RANs) dominate energy expenditures, accounting for over 50\% of  total network energy consumption \cite{nextgen}.
When transitioning to the 5$^{\text{th}}$ generation (5G) RAN, many energy-hungry technologies such as millimeter wave (mmWave) and massive multiple-input multiple-output (mMIMO), along with novel energy-efficiency techniques such as mMIMO muting, lean-carrier design, sleep modes and machine learning based methods \cite{lopez2022survey}, have been introduced. This is directly related to the evolution of users' needs, which have evolved from voice communications (generating traffic as low as 100 kbps) to heterogeneous and bandwidth-greedy applications, such as gaming, streaming, and virtual/augmented reality (VR/AR), that require data rates of up to 250 Mbps. Moreover, users are frequently moving, which causes the demand to spatially and temporally shift across the network. In particular, the proliferation of mobile Internet-of-things (IoT) devices (e.g., unmanned aerial vehicles [UAVs] and autonomous vehicles) contribute to this phenomenon. 

\begin{figure*}[ht]
\centering
\includegraphics[width=0.85\linewidth]{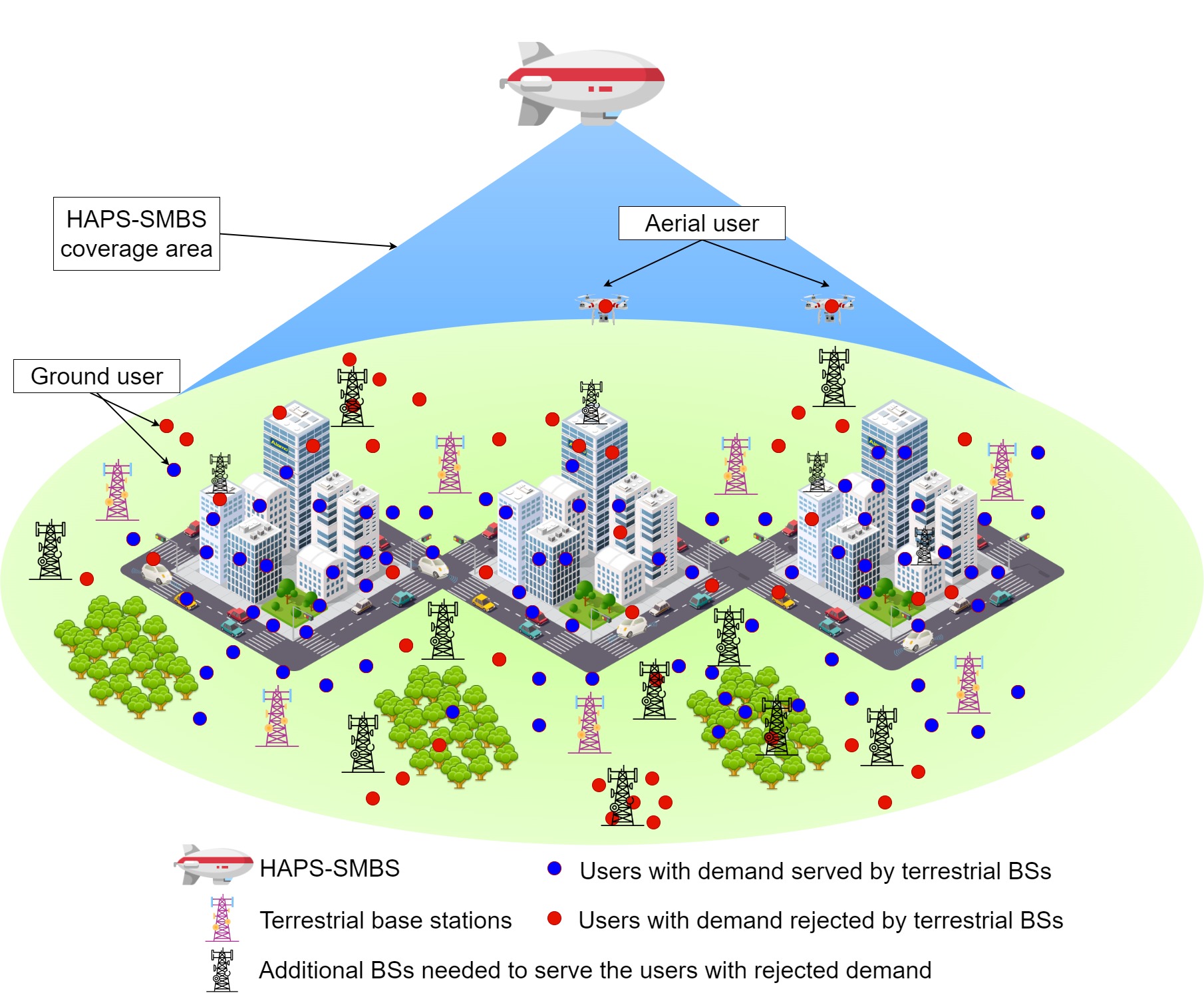}
\caption{A HAPS-SMBS used as a complement to an existing RAN serving the mobile demand that is rejected by the terrestrial BSs due to random spatial and temporal changes in demand.}
\label{fig:1}
\end{figure*}

To avoid service outages, mobile network operators (MNOs) tend to over-provision by densifying the network with more base stations (BSs), and these BSs are clustered in areas that may experience peak demands, such as urban areas. However, this approach degrades the network's energy efficiency due to the inefficient use of BSs. Although several energy-saving techniques exist to mitigate this problem \cite{lopez2022survey}, densification still increases the capital and operational expenditures (CapEx and OpEx) of terrestrial networks. Consequently, densification is unsustainable in the long run. 

To tackle this issue, we propose the use of high altitude platform station (HAPS) nodes.
We envision the use of HAPS as a complement to existing RANs for their capacity to respond to surges in terrestrial network traffic, as depicted in Fig. \ref{fig:1}. Thanks to recent advancements in avionics, mMIMO, battery and solar energy, HAPS are emerging as a main component of beyond 5G networks \cite{kurt2020vision,alam2020high,Jaafar2022}. When used as a super macro base station (SMBS), a HAPS can provide reliable wireless connectivity through strong line-of-sight (LoS) links in areas up to 500 km in radius. Through mMIMO and advanced multiple access techniques, such as precise beamforming and steering, HAPS can respond to rapid temporal and spatial changes in traffic demand. 
In light of these advantages, a HAPS can be seen as an energy-efficient and cost-saving substitute for dozens or even hundreds of terrestrial BSs in dynamic and unpredictable peak demand areas.

\section{Terrestrial Cellular Networks: Current Challenges and Solutions}

\subsection{Ubiquitous and Ultra-Reliable Connectivity}

Traffic demand in wireless networks has evolved substantially in the last ten years because of new applications with stringent QoS requirements. For instance, autonomous vehicles need full coverage and 99.999\% reliable communication links \cite{Jaafar2022}, while VR/AR applications require high throughput of up to 250 Mbps. In addition, given the massive deployment of fixed and mobile IoT devices (e.g., in smart cities), networks should be able to handle the ultra-dense traffic generated by these devices and their mobility. Such constraints call for greater network agility to satisfy this heterogeneous, dynamic, and unpredictable massive traffic.

To effectively address this issue in cellular networks, it needs to be tackled from the perspectives of backhaul and RAN. From the backhaul perspective, novel technologies such as network slicing, software-defined networks, and virtualization have been efficiently developed and adopted in 3GPP Release 17. However, from the RAN perspective, research is still ongoing to define an effective structure to meet the ubiquitous and reliable connectivity requirements of future wireless networks.

Among the most popular solutions of MNOs in coping with dynamic and unpredictable ultra-dense traffic is \textit{network densification}. As today's wireless infrastructure is cell-based using BSs, the only way to guarantee ubiquitous coverage and handle traffic spikes is to carefully add new BSs. Hence, the cellular network turns into a heterogeneous network (HetNet), which consists of cells of various sizes (e.g., macrocells and small cells).
In HetNets, a number of small cells are added within the coverage areas of macro cells to enhance the network's capacity and handle the peak traffic demands. However, since the mobile traffic demand is becoming more dynamic and unpredictable, it may spike at random times and locations. Therefore, the extra capacity provided by the small cells may be useful at certain times and areas, but underutilized for the remaining time. Although these additional cells can be switched off to save energy when unused, this approach may increase the complexity of the network's planning, since the resource allocation strategy needs to be continuously adjusted to cope with the cell switching on-off dynamicity. 
Alternatively, transportable base stations can be used to serve predictable demand peaks, e.g. during concerts and sports events, or to provide temporary connectivity in disaster areas. However, they are not agile enough to cover the randomly occurring demand peaks, not to mention the energy and manual efforts needed to transport them constantly. Consequently, densifying the network with additional cells is the common solution to the ubiquitous connectivity problem.

\subsection{Efficient Energy Consumption} Despite the efforts to design 4G and 5G networks in an energy-efficient manner, their energy consumption remains high as the RAN equipment consumes power almost constantly. If the same policies are kept in place, RAN energy consumption is expected to exceed the 100 Terawatt-hour (TWh) mark by 2030 (from the current mark of 77 TWh), and its carbon footprint is expected to double \cite{Yuan2021}.


To overcome the high energy consumption, \textit{cell switch-off} and \textit{sleeping} methods are being proposed. Cell switch-off involves completely shutting down a cell if, for instance, traffic in the area can be handled by neighboring cells. Sleeping involves deactivating certain hardware components of a BS for short periods of time and reactivating it according to the absence or presence of traffic. However, these approaches may not be the most practical. If a cell is switched-off and then reactivated, it can take from a few seconds to a few minutes to re-start it and configure it, which could be too long if urgent services were required \cite{lopez2022survey}. Also, in sleep mode, the BS's energy consumption is not null, but rather reduced compared to its active mode. In addition, these methods make network management more complex since dynamic and adaptive resource allocation must be handled for an uninterrupted service. Although AI-native solutions can be called in for traffic forecasting and managing sleep modes, they require significant computational power and storage capabilities, which can undermine the intended energy gain \cite{lopez2022survey}.

\section{Supporting Dynamic and Unpredictable Traffic Demand with HAPS}

The idea of using HAPS-SMBS instead of BS-based network densification is motivated by the unique properties of HAPS. First, HAPS operate at an altitude of 20 km to provide a wide coverage area, which is evaluated to be 35 km in radius at an elevation angle of 30$^{\circ}$. Consequently, a single HAPS can substitute for dozens of terrestrial BSs to provide an equivalent coverage footprint. Therefore, when traffic is dynamic and unpredictable over a wide area, a HAPS can handle it in an efficient manner when compared to BS densification.

HAPS also benefit from a high payload capacity, which allows them to rely on solar energy and batteries to fly for months or even years without interruption \cite{kurt2020vision}. This is important, as HAPS operation can achieve net zero carbon emission, in contrast to terrestrial wireless networks.
Moreover, HAPS can host different types of payloads, such as for communication, computing, and storage, in order to support data center-like services.

Although HAPS operate at altitudes of around 20 km, communication links to and from ground users enjoy a strong line-of-sight (LoS) component. That is, users can directly ``see'' a HAPS, as opposed to communications with BSs where links are often obstructed by buildings, trees, or vehicles. The strong LoS enables a high signal-to-noise ratio (SNR), which, despite the long distance between HAPS and ground users, is equivalent to that between terrestrial BSs and users. For instance; given the same frequency, a path-loss exponent of 2 for a HAPS-user link,
and a path-loss exponent of 4 for a terrestrial BS-user link in a dense urban area,
the received signal strength would be the same for the BS-user communication at a distance of 141 m and for the HAPS-user communication
at a distance of 20 km. Moreover, such a link achieves a communication latency of 0.067 ms, which is below that of low-Earth-orbit (LEO) satellites at 300 km altitude, evaluated at 1 ms, and sufficiently low to support uRLLC-like services \cite{alam2020high}.

When compared to LEO satellites, another potential complementary to terrestrial mobile networks, HAPS present further advantages in terms of sustainability. First, due to its quasi-stationary location, a single HAPS can be used to cover a specific area continuously, while a constellation of at least three LEO satellites is needed to serve the same area, due to the latter's fast orbital motion (which also results in frequent handovers). Secondly, although the production and launch costs of LEO satellites are steadily decreasing, a rocket launch is required for their deployment, which can be a cumbersome and non-environmentally-friendly process. In contrast, HAPS can be deployed by means of aviation, or simply based on buoyancy. Finally, HAPS can be recovered, upgraded, and reused during or after its service period, while LEO satellites might end up as space junk for years until falling out of orbit and burning up in the atmosphere.

Alternatively, one could argue that instead of using a HAPS, an SMBS with the same capacity can be installed as a terrestrial BS, which would enhance the capacity through the addition of several sectors to the BS. However, as the number of sectors increases, their serving apertures become narrower. This could generate frequent handoffs for mobile users that are close to the terrestrial SMBS, thus degrading the quality of  the communication and adding unnecessary overhead. By contrast, a HAPS-SMBS enjoys a bird's-eye view with wider sector areas enabled by cylindrical mMIMO antennas \cite{Tashiro2021}. 
Combined with strong LoS channels, the use of a HAPS-SMBS emerges as a more attractive connectivity solution than its terrestrial counterpart.

\section{Case Study}
In order to demonstrate the efficiency of a HAPS-SMBS in handling dynamic and unpredictable traffic demands, compared to the conventional method of network densification, we present a case study simulation. Our system consists of an $8\times 8$ km$^2$ urban area (e.g., a university campus or an urban residential neighborhood), where 14,000 users are distributed uniformly and served by 36 initial BSs that are placed uniformly. We assume that each BS has a coverage area of 700 m in radius and supports a RAN capacity of 1 Gbps in accordance with IMT-2020 specifications. Please note that all BSs are assumed to have the same capacity and coverage specifications, therefore all BSs are considered macro BSs. We assume that time is divided into equal time slots (TSs), such that at each TS the user demands are defined as a throughput requirement in Mbps, and are assumed to follow a half-normal distribution $|\mathcal{N}(0,20)|$, which corresponds to a mean demand of 16 Mbps. For the sake of simplicity, we assume that a user's demand can be served by a BS as long as the user is within the coverage footprint of the BS and that the sum of served demands does not exceed the BS's capacity. Finally, users and their demands are assumed to be fixed for one TS, but can vary from one TS to another, where the random waypoint mobility model is adopted in our simulations.

\begin{figure}[t!]
    \centering
    \includegraphics[width=0.9\linewidth]{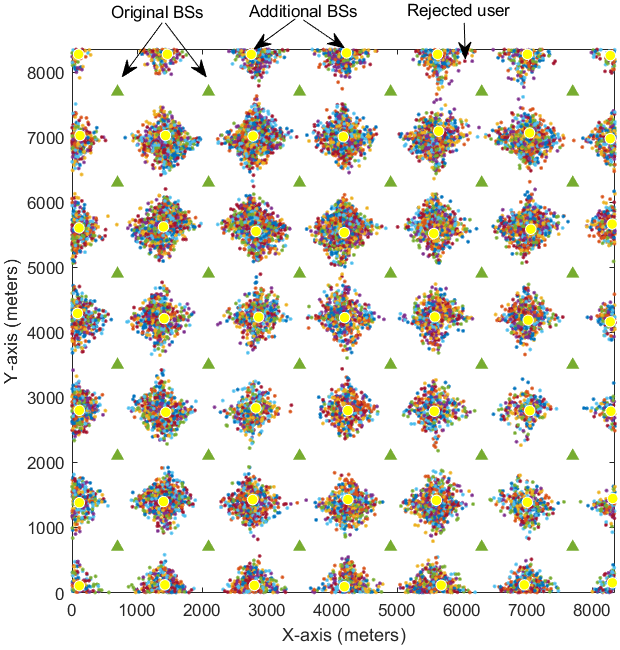}
        \caption{Network densification showing original BSs, rejected users in time, and additional BSs to support these users.}
    \label{fig:2}
\end{figure}

First, in the case of a single TS where all system parameters are fixed (i.e., user locations and traffic demands) the initial 36 BSs are expected to be sufficient to handle all traffic demands. However, in a real-world scenario, users will move around and behave differently in terms of service demand. To illustrate this case, we simulate the system with changing user locations and demand every TS=1 minute, for a period of 24 hours (i.e., for 1,440 TSs). 
Without loss of generality, we assume that each user is associated with the closest BS and that the latter rejects out-of-coverage users and/or until the sum of traffic demand is equal to its RAN capacity. As shown in Fig. \ref{fig:2}, the initial BSs would reject users at some TSs as their capacity would be exceeded by the presence of several users and high demand in its coverage area (colored dots correspond to different users/TSs). In this example, an average of 1\% of users are rejected, while the network's \textit{capacity utilization}, defined as the ratio of the served traffic demand to the total RAN capacity of the network, is 73\%.

To recover rejected users, conventional network densification would cluster them and deploy additional BSs. As illustrated in Fig. \ref{fig:2}, 49 additional BSs (yellow dots) are needed to connect 100\% of the users. However, the corresponding network's capacity utilization (i.e., for the initial 36 BSs and additional 49 BSs) is reduced to 31.1\% due to the under-utilization of the additional BSs over time, as shown in Table \ref{tab:table1}.
Even by switching off the BSs when they are not used, the network's capacity utilization improves slightly to 34\%.

Alternatively, assuming that a single HAPS-SMBS is deployed above the area of interest with a capacity of 2 Gbps, all of the rejected demands can be handled. For this scenario, the related network's capacity utilization (i.e., for the 36 initial BSs and 2 Gbps HAPS) is 71.2\%, which is very high compared to the network densification benchmark. In addition, the HAPS-assisted system deploys a smaller communication payload than network densification, which results in a lower power consumption of 140.6 kW.

\begin{table}[t]
\centering
\caption{Comparison of HAPS-assisted network and network densification  approaches (Avg. demand per user is 16 Mbps)}
\label{tab:table1}
\renewcommand{\arraystretch}{2}
\begin{tabular}
{p{0.2\linewidth}p{0.38\linewidth}p{0.26\linewidth}}
\hline
\textbf{Feature} & \makecell[l]{\textbf{HAPS-assisted} \\ \textbf{Network}} & \makecell[l]{\textbf{Densified }\\ \textbf{Terrestrial Network}}\\
\hline
\hline
\makecell[l]{Available\\ Capacity} & \makecell[l]{38 Gbps (36 Gbps terrestrial \\+ 2 Gbps HAPS)} & \makecell[l]{85 Gbps} \\
\hline
\makecell[l]{Proportion of\\ Users Served} & 100\% & 100\% \\
\hline
\makecell[l]{Capacity \\Utilization} & \makecell[l]{71.2\%} & \makecell[l]{31.3\%} \\
\hline
\makecell[l]{Network Power\\ Consumption}
& \makecell[l]{140.6 kW} & \makecell[l]{314.5 kW} \\
\hline
\end{tabular}
\end{table}

\subsection{Impact of Average Demand}

\begin{figure}[t]
\centering
\includegraphics[width=1.08\linewidth]{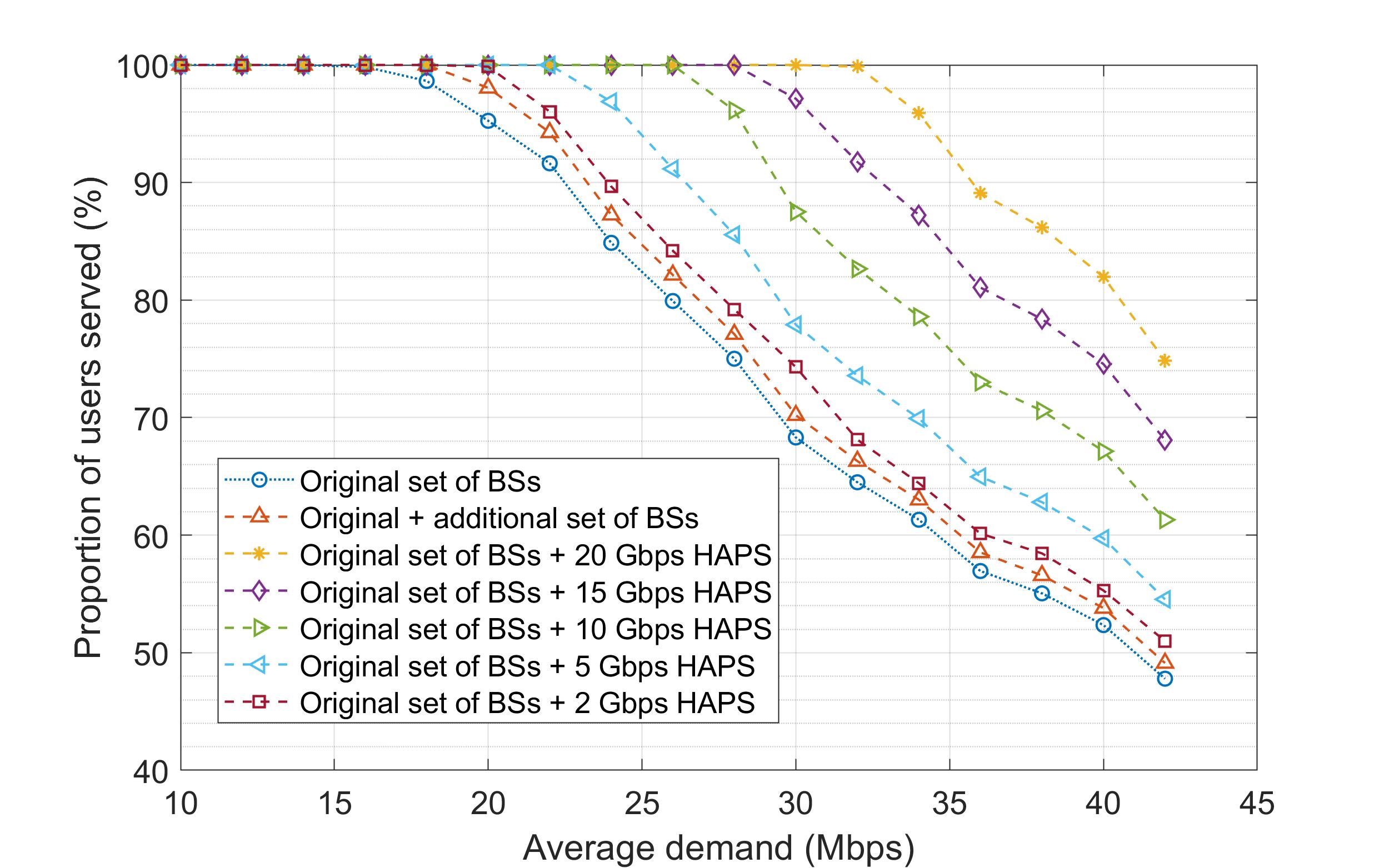}
\caption{Proportion of users served vs. average demand per user.}
\label{fig:PvD}
\end{figure}

\begin{figure}[t]
\centering
\includegraphics[width=1.08\linewidth]{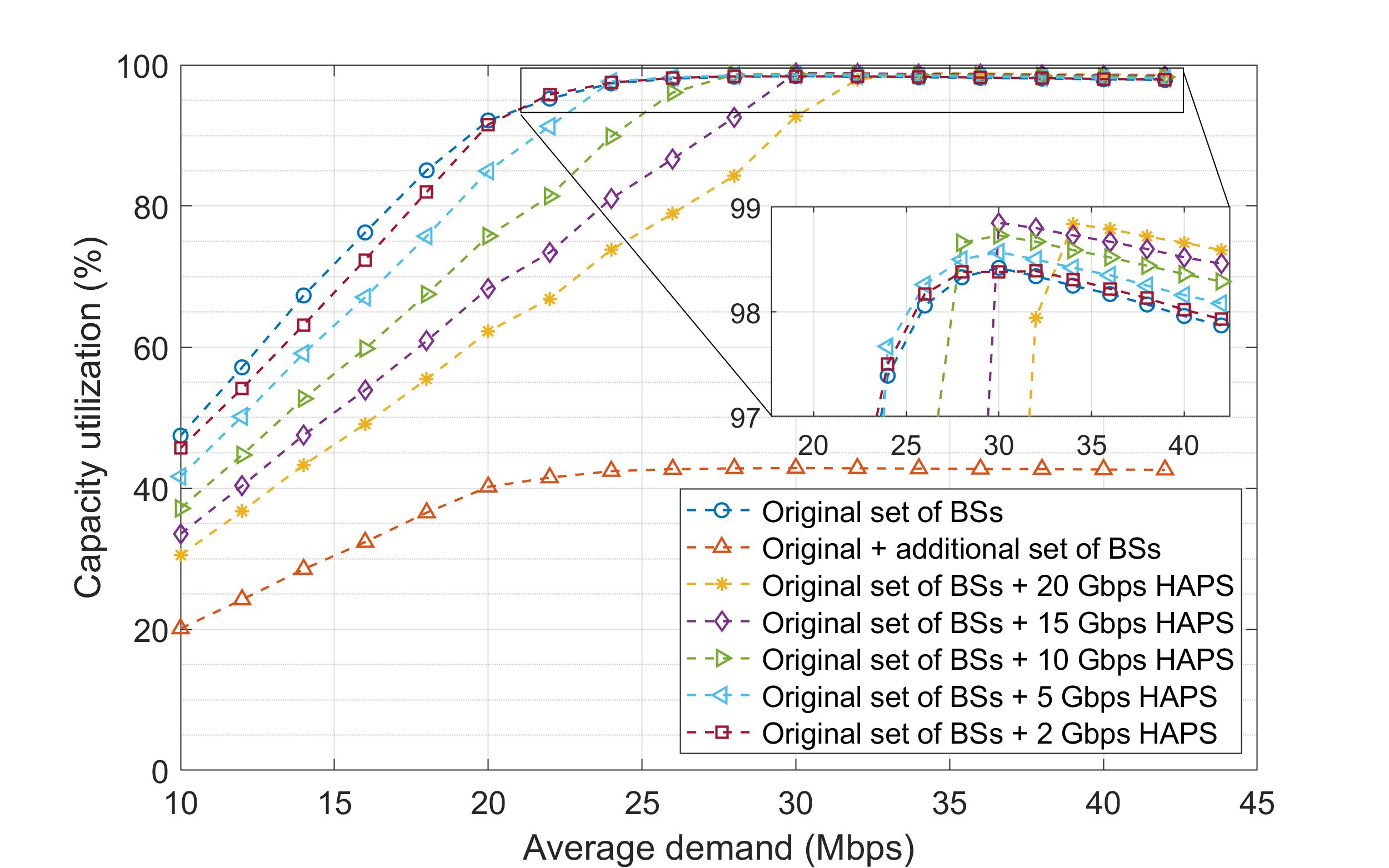}
\caption{Capacity utilization vs. average demand per user.}
\label{fig:UvD}
\end{figure}

Here we examine the impact of increasing the traffic demand on the original network (``Original set of BSs''), conventional network densification (``Original + additional set of BSs''), and HAPS-assisted network (``Original set of BSs + $X$ Gbps HAPS''), where $X \in \{ 2,5,10,15,20 \}$. Related results are presented in Figs. \ref{fig:PvD} and \ref{fig:UvD}, where the former depicts the proportion of users served and the latter shows the capacity utilization, both averaged over time and as functions of the average demand per user.

According to Fig. \ref{fig:PvD}, for any system considered, the proportion of users served is at 100\% until a critical average demand from which it starts dropping. For instance, the performance of the ``Original set of BSs'' starts dropping from an average demand of 14 Mbps, while for ``Original + additional set of BSs'', the performance degrades from an average demand of 16 Mbps. By contrast, the HAPS-assisted network's performance not only tolerates a higher critical average demand value (between 20 and 32 Mbps), but also the proportion of users served remains higher (from 2\% for 2 Gbps HAPS up to 26\% for 20 Gbps HAPS at the average demand of 42 Mbps) compared to the densified network. This is thanks to the large coverage area of the HAPS, which can serve users regardless of their location as long as its capacity is not exceeded. These results indicate the robustness of the HAPS-assisted network in dealing with the increasing dynamic traffic demands of users.

Based on the plots of Fig. \ref{fig:UvD}, as the average demand increases, the performance in terms of capacity utilization for all approaches demonstrates a two-phase behavior. At first, capacity utilization increases in proportion to average demand until reaching a maximal value point (phase 1). Beyond this point, capacity utilization starts to slowly decrease (phase 2).
In phase 1, the higher the average demand value, the more capacity is used from the network. Saturation is reached at the critical point, which means that the network is used at full capacity to serve the percentage of users within reach. In phase 2, the capacity utilization slightly degrades, since higher average demand forces the network to drop users who now have higher demand.

We notice that both the ``Original set of BSs'' and ``Original + additional set of BSs'' have the same critical point at an average demand of 30 Mbps. However, the HAPS-assisted system demonstrates different critical points, depending on the HAPS's capacity. In fact, the critical average demand value is higher when more capacity is available at the HAPS. Moreover, its phase 1 capacity utilization is the highest when the HAPS is equipped with the smallest capacity (before reaching its critical point), while it is the lowest in phase 2. Nevertheless, in phase 2, the capacity utilization gaps for different HAPS capacities are very small (up to 2\%).  This means, depending on what the average demand in the system would be, it is recommended for the HAPS to be equipped with several BSs; however, it would switch on the smallest number of BSs that guarantee the highest capacity utilization in phase 1. Finally, the 2 Gbps HAPS solution outperforms the ``Original + additional set of BSs'' in terms of capacity utilization while achieving a very close performance to the ``Original set of BSs''. This is an expected result since the HAPS-assisted network can serve a higher number of users than the densified network while introducing less amount of capacity to the system.

\subsection{Sustainability Analysis}
Here, we compare terrestrial networks and HAPS-SMBS systems from the three perspectives of sustainability, which is summarized in Table \ref{tab:table2}. 

\subsubsection{Economical analysis}
The CapEx of terrestrial BSs consists of the RAN equipment, site buildout, and site installation costs. Adding to that, the OpEx of such BSs is composed of operation and management (O\&M), site lease, line lease, and transmission costs. On the other hand, a HAPS-SMBS consists of a flight control subsystem, an energy management subsystem, and a communications payload subsystem \cite{kurt2020vision}. When compared with terrestrial BSs, a HAPS-SMBS is free of the site lease, site buildout and site installation costs. Furthermore, a HAPS-SMBS uses a wireless backhaul link to its ground station for routing the traffic. Therefore, it requires less ground infrastructure when compared to terrestrial BSs. Thanks to the wide coverage area of HAPS, more than one HAPS can use the same ground station, which would reduce the ground infrastructure needed even further. 

Although HAPS-SMBS requires less ground infrastructure and lease, the aircraft cost increases the CapEx and OpEx of this option. The costs of building a HAPS aircraft as well as terrestrial BSs are trade secrets, hence making a direct comparison is very difficult. Nevertheless, in our case the HAPS-SMBS can have the same CapEx and OpEx of 49 terrestrial BSs, which are significantly large amounts, for the two approaches to be economically comparable. Note that, in our case we are considering an 8 km by 8 km area, whereas a HAPS-SMBS coverage area can be much bigger depending on the scenario. Hence, the CapEx and OpEx of HAPS-SMBS solution can be less than that of network densification depending on the number of terrestrial BSs the HAPS-SMBS is substituting. For instance, in the case of Denmark, 8 HAPS with 10 Gbps capacity each can be used instead of rolling out 1261 (resp. 1810) BSs in the worst (resp. likely) case \cite{gsma2022}; which corresponds to  a range of 157 to 226 base stations per HAPS. The total cost of such a terrestrial rollout ranges from 300M\EUR{} to 530M\EUR{} corresponding to an economically viable cost between 37.5M\EUR{} and 66.25M\EUR{} per HAPS. Note that the lifetime of a HAPS is expected to be between 8 and 10 years \cite{gsma2022}, which is comparable to the lifetime of a 5G BS, estimated between 5 and 10 years \cite{nokia}.

\begin{table}[!t]
\renewcommand{\arraystretch}{2}
\caption{Comparison of the sustainability of HAPS-SMBS and terrestrial networks}
\label{tab:table2}
\centering
\begin{tabular}{p{0.22\linewidth}p{0.05\linewidth}p{0.05\linewidth}p{0.05\linewidth}p{0.09\linewidth}p{0.14\linewidth}}
\hline
\makecell[l]{\textbf{Feature}} & \multicolumn{3}{c}{
{\textbf{Sustainability} \textbf{Aspect}}} & {\textbf{HAPS}\textbf{-SMBS}} & {\textbf{Terrestrial} \textbf{networks}}\\
\hline
\hline
& \textbf{Eco.} & \textbf{Env.} & \textbf{Soc.} && \\
\hline
\makecell[l]{\textbf{Ground space} \\\textbf{required}} & \ding{51} & \ding{51} & -- & No & Yes \\
\hline
\makecell[l]{\textbf{Renewable energy} \\\textbf{penetration}} & \makecell[l]{\ding{51}} & \makecell[l]{\ding{51}} & \makecell[l]{--} & \makecell[l]{High} & \makecell[l]{Low} \\ 
\hline
\makecell[l]{\textbf{Cabling} \\\textbf{infrastructure}}  & \makecell[l]{\ding{51}} & \makecell[l]{\ding{51}} & \makecell[l]{--} & \makecell[l]{Low} & \makecell[l]{High} \\
\hline
\makecell[l]{\textbf{Visual} \\ \textbf{pollution}} & -- & \makecell[l]{\ding{51}} & \makecell[l]{\ding{51}} & \makecell[l]{Very\\ Low} & \makecell[l]{High} \\ \hline
\makecell[l]{\textbf{RF} \\\textbf{exposure}} & -- & \makecell[l]{\ding{51}} & \makecell[l]{\ding{51}} & \makecell[l]{Very\\ Low} & \makecell[l]{High} \\
\hline
\end{tabular}
\end{table}

\subsubsection{Environmental analysis}
HAPS can operate with less carbon emission than a terrestrial network since it mainly relies on renewable energy to operate (e.g., solar energy).
Given the advancement in PV technology, it is expected that HAPS systems will be able to operate fully with energy produced by its PV panels \cite{renga2022can}. Although terrestrial BSs can also be supported with renewable energy, physical and environmental restrictions reduce the efficiency of such energy-harvesting mechanisms for them. For instance, terrestrial BSs do not usually have large areas where PV panels or wind turbines can be installed, especially in urban areas where real estate is much more expensive. Moreover, terrestrial BSs can receive less solar radiation than HAPS due to atmospheric attenuation.

Deploying more terrestrial BSs requires additional infrastructure to be installed such as power lines and fiber-optic links, which can impact the environment as they may alter natural ecosystems (e.g., lakes and forests) and increase the carbon footprint of a terrestrial network. By contrast, a HAPS can leverage its strong LoS communication capability and hybrid backhauling, using radio-frequency (RF) and free space optics (FSO), to provide communication services without any additional ground construction or disturbance to the ecosystem. 

\subsubsection{Societal analysis}

The crowding of terrestrial BSs in densely populated areas can make residents feel uncomfortable. 
From a health perspective, although clinical results about adverse effects from close RF exposure to a BS were inconclusive \cite{roosli2010systematic}, symptoms reported by people living near BSs have included fatigue, sleep disturbances, and headaches. This has led to resistance against new deployments of BSs, and it has affected the market value of houses close to BSs. 
Using a HAPS as an SMBS over an urban area can ease this problem since the RF source would be sufficiently far and invisible.
From a cityscape perspective, there are concerns that network densification with BSs would visually pollute the cityscape, and this concern is particularly pronounced in touristic spots. Camouflaging BSs as street lamps and trees is a common workaround; however, it is not a universal remedy. Instead, a HAPS can be exploited to support the traffic demand in these areas while preserving the beauty and reputation of the city.

On the other hand; legal, aviation and regional regulations may complicate the rollout of HAPS-SMBS systems. First, there is an ongoing dispute regarding where the airspace ends and outer-space begins. This is important for HAPS since it would define which laws will be applied to HAPS aircraft. Note that each country has the right to apply its own laws and regulations over its airspace. However, countries do not have any jurisdiction in outer space above their territory.
In addition, although HAPS systems fly above the air control limits, launching and recovering HAPS may require coordination with airspace control agencies. Moreover, the aviation requirements may vary with the type of HAPS, e.g., unmanned balloons, gliders, or airships, as well as with the aviation authority.
Also, some of the frequencies allocated by the International Telecommunication Union (ITU) for HAPS systems are not globally assigned, which could complicate the standardization of HAPS \cite{kurt2020vision}.

\section{Challenges and Future Research Directions}
Leveraging a HAPS as a complement to a terrestrial RAN can be a sustainable solution to the issue of ever-increasing and dynamic traffic demand. However, to realize the full potential of this solution, some challenges need to be addressed. We refer the interested readers to \cite{kurt2020vision} for a comprehensive overview on these issues. Note that there are many proposals to solve these issues in the field of integrated satellite-terrestrial networks \cite{rinaldi2020non}, which could be utilized to find the solutions.

\subsection{Mobility and Vertical Handoff Management}
 Due to the mobility of users, frequent handovers in HAPS-assisted networks are expected among terrestrial BSs (horizontal handoff) and between the HAPS and BSs (vertical handoff). Although managing horizontal handoffs is a classical problem, managing vertical handoffs is a novel issue that needs to be carefully tackled. Indeed, the overlap of HAPS and terrestrial BS coverage areas may favor frequent and undesired handoffs between the HAPS and terrestrial BSs due to the strong LoS link to/from HAPS.

 \subsection{Radio Resources and Interference Management}
 The overlap of HAPS and terrestrial BS coverage areas is likely to experience high interference, due in part to the large coverage area, strong LoS links, imprecise signal directivity, and radio resources reuse by HAPS. Indeed, how to effectively allocate the radio resources to minimize interference at ground BSs and users is a key issue, especially when the same frequency bands are used by both HAPS and terrestrial systems. To solve this problem, integrated HAPS-terrestrial network solutions need to be developed, where resources are seamlessly and smartly shared between HAPS-enabled and terrestrial BSs-enabled communications. 

\subsection{Energy Harvesting and Management}
One of the major bottlenecks of HAPS is its limited energy supply. Many solutions ranging from hydrogen cells to laser charging are being discussed to sustain HAPS energy consumption. Currently, the most practical solution appears to be PV panels combined with Li-ion batteries. Although the latter's costs are decreasing and their efficiencies are increasing, they need to support more energy-hungry communication technologies, such as mMIMO and mmWave. In addition, existing methods for network energy preservation, such as power control and cell switch-off, can be extended to HAPS. Furthermore, wireless power transfer (WPT) and laser charging techniques can be investigated for use with HAPS-enabled systems.
Novel technologies such as reconfigurable intelligent surfaces (RISs) can be leveraged to improve WPT quality.

\section{Conclusion}
In this article, we discussed the potential of HAPS-SMBSs to complement terrestrial networks in supporting the massive, dynamic, and unpredictable traffic demands of users. First, we identified the current challenges and existing solutions of terrestrial networks in handling dense and dynamic traffic. Then, we proposed HAPS-SMBS as an alternative to network densification. A case study was developed to quantify the performance of a HAPS-SMBS in terms of the proportion of users served and capacity utilization in highly dynamic traffic environments. The case study showed that the HAPS-assisted terrestrial network can sustain the demand with less power consumption and more efficiency in terms of capacity utilization, when compared to the densified terrestrial network. Through a sustainability analysis, we examined the potential economic, environmental, and social impacts of a HAPS-SMBS. We also drew attention to several challenges that need to be addressed in order to unleash the full potential of HAPS-enabled systems.



\bibliographystyle{IEEEtran}
\bibliography{paper.bib}

\begin{IEEEbiographynophoto}{Cihan Emre Kement} [M]
(cihankement@sce.carleton.ca) is a Postdoctoral Fellow at Carleton University, ON, Canada. His research interests include cyber-physical systems, wireless communications and optimization.
\end{IEEEbiographynophoto}
\vspace{-1\baselineskip}
\begin{IEEEbiographynophoto}{Ferdi Kara} [SM] (f.kara@beun.edu.tr) is an Assistant Professor at Zonguldak Bulent Ecevit University, Turkey. He is also a Postdoctoral Fellow at Carleton University, ON, Canada. His research interests are PHY layer aspects of wireless communications, and ML applications.   
\end{IEEEbiographynophoto}

\vspace{-1\baselineskip}
\begin{IEEEbiographynophoto}{Wael Jaafar}
[SM] (wael.jaafar@etsmtl.ca) is an Associate Professor at École de Technologie Supérieure, QC,
Canada. His research interests include aerial networks, 5G and
beyond technologies, and machine learning.
\end{IEEEbiographynophoto}

\vspace{-1\baselineskip}
\begin{IEEEbiographynophoto}{Halim Yanikomeroglu}  [F] (halim@sce.carleton.ca) is a Full Professor at Carleton University, ON, Canada. His current research focus is non-terrestrial networks (NTN) for the 6G era. 
\end{IEEEbiographynophoto}

\vspace{-1\baselineskip}
\begin{IEEEbiographynophoto}{Gamini Senarath}  [M] (gamini.senarath@huawei.com) is a Distinguished Member of Staff at Huawei Technologies, Canada. His current research interests include software defined networks, virtual network embedding, resource management of third party infrastructure for 5G wireless systems.
\end{IEEEbiographynophoto}

\vspace{-1\baselineskip}
\begin{IEEEbiographynophoto}{Ngọc Dũng Đào} (ngoc.dao@huawei.com) is a principle engineer at Huawei Technologies, Canada. His recent research interests include 5G and beyond mobile network architectures, data analytics, vehicle communications, IoT, and multimedia.
\end{IEEEbiographynophoto}

\vspace{-1\baselineskip}
\begin{IEEEbiographynophoto}{Peiying Zhu} [F] (peiying.zhu@huawei.com) is a Huawei Fellow. She is currently leading 5G wireless system research at Huawei Technologies Canada. The focus of her research is advanced wireless access technologies with more than 150 granted patents. 
\end{IEEEbiographynophoto}




\end{document}